\documentclass[%
reprint,
superscriptaddress,
 amsmath,amssymb,
 aps,
]{revtex4-2}

\usepackage{graphicx}
\usepackage{dcolumn}
\usepackage{bm}

\begin{document}

\preprint{APS/123-QED}

\title{Invertible Optical Nonlinearity in Epsilon-near-zero Materials}

\author{Chentao Li}
 \affiliation{Department of Physics, Emory University, 400 Dowman Dr., Atlanta, Georgia 30324, United States}
\author{Xinyu Tian}%
 \affiliation{Department of Physics, Emory University, 400 Dowman Dr., Atlanta, Georgia 30324, United States}
\author{Guoce Yang}
 \affiliation{Department of Physics, Emory University, 400 Dowman Dr., Atlanta, Georgia 30324, United States}%
\author{Sukrith U. Dev}
 \affiliation{Air Force Research Laboratory, Munitions Directorate, Eglin AFB, Florida 32542, United States}
 \author{Monica S. Allen}
 \affiliation{Air Force Research Laboratory, Munitions Directorate, Eglin AFB, Florida 32542, United States}
\author{Jeffery W. Allen}
 \affiliation{Air Force Research Laboratory, Munitions Directorate, Eglin AFB, Florida 32542, United States}
\author{Hayk Harutyunyan}
 \email{hayk.harutyunyan@emory.edu}
 \affiliation{Department of Physics, Emory University, 400 Dowman Dr., Atlanta, Georgia 30324, United States}

\date{\today}

\begin{abstract}
Epsilon-near-zero (ENZ) materials such as indium tin oxide (ITO), have recently emerged as a new platform to enhance optical nonlinearities. Here we report a theoretical and experimental study on the origin of nonlinearities in ITO thin films that are dominated by two mechanisms based on intraband and interband transitions. We show that there are two competing factors that jointly contribute to a spectrally-invertible nonlinearity of ITO near its ENZ region i.e. the non-parabolicity of the band structure that results in a larger effective mass in the intraband transition and the Fermi energy shift, which determines the free carrier density. Our work reveals the relationship between the large nonlinearity and the intrinsic material properties of the ITO films.
\end{abstract}

\maketitle



The pursuit of a large optical nonlinearity with a moderate excitation field has been a long-time goal in the study of light-matter interactions. Most approaches that focus on significant enhancement of nonlinearities are based on materials whose refractive index can be drastically modified using an optical pump ~\cite{RN32,RN56}. Recently, epsilon-near-zero (ENZ) materials have attracted considerable attention due to their intriguing optical properties, such as diverging velocities~\cite{li2015chip,vesseur2013experimental}, wavelength expansion~\cite{reshef2017direct} and large field enhancement~\cite{boyd,RN42,RN10}. While the pioneering work focused on realization of ENZ and its linear properties~\cite{RN43,RN22}, recent  studies have revealed the great potential of ENZ materials in nonlinear applications~\cite{RN35}. An intuitive calculation shows that the change of refractive index can be expressed as $dn=d\epsilon/2\sqrt{\epsilon}$, which becomes significant given a near-zero permittivity $\epsilon\approx0$.

Transparent conductive oxides (TCOs), a subset of ENZ materials, have emerged as a new platform to enhance optical nonlinearities ~\cite{RN20,luk2015enhanced,capretti2015enhanced,RN19}. Typical examples that have been widely investigated include Al or Ga doped zinc oxides (AZO/GZO)~\cite{RN20,RN37,allen2013application}, cadmium oxides (CdO)~\cite{RN9,RN23} and ITO~\cite{RN30,RN33}. These materials support ENZ resonances in the near-infrared (NIR) range and are compatible with existing complementary metal-oxide semiconductor (CMOS) technologies~\cite{babicheva2013towards,babicheva2015transparent}, thus show great promise for applications in integrated telecom industries~\cite{liberal2017rise}. Moreover, the stability of these materials is better than traditional metals~\cite{johnson1974optical} and highly-doped semiconductors~\cite{RN54} that show near-zero $\epsilon$ at their bulk plasma wavelengths, and therefore can endure larger optical pump fields that can be used to induce larger nonlinearities without reaching the damage threshold. This rich design space for photon manipulation enables different strategies to enhance the nonlinearity, such as the use of patterned structures ~\cite{RN29,RN52}, coupling with plasmonic materials ~\cite{RN30,RN11,RN26,RN31} and utilizing waveguide modes~\cite{RN21}.

Recent studies have demonstrated the possibility of exploiting transient transmittance (reflectance) spectroscopy to study the temporal dynamics of large nonlinearities~\cite{RN7,RN10,RN30,RN19,RN37}. A significant change of refractive index $dn$ on the order of unity was observed in transient experiments conducted on ITO thin films~\cite{boyd,RN57}. The ultrafast transient response in these materials is dominated by two sub-picosecond (ps) processes: (i) Intraband transition excited by pump photons with sub-bandgap energies; and (ii) Interband transition, pumped with high energy photons such as UV region photons. Understanding the origin of the nonlinearity observed in ultrafast experiments necessitates effective modelling of these two processes~\cite{RN30,boyd,RN33}. Interband transition is described by carrier generation followed by trap-assisted recombination~\cite{RN10}. For intraband transitions, an optical Kerr effect model was proposed, where the plasma frequency $\omega_p$ or the refractive index $n$ is linearly dependent on the pump intensity, enabling reasonable quantitative description of the nonlinearity~\cite{RN19,RN5}. Specifically, the two-temperature model reveals that the rise of the electron temperature due to laser excitation generates temperature-dependent complex index~\cite{boyd,RN30,RN10}. This approach uses fitting parameters, such as electron-phonon coupling strength G determined from time-resolved index change measurements $\Delta{n(t)}$. Alternatively, a deterministic (fitting parameter free) physical model attributed the nonlinearity to the non-parabolic conduction band structure, which greatly modifies the electron effective mass for the excited hot electron population~\cite{RN1,RN23,RN41,RN52}. However, to our knowledge, no unified approach exists to date that describes the excited state dynamics and instead existing research models these mechanisms as separate processes when explaining the origin of the nonlinearity, while their joint contributions and simultaneous interactions remain unexplored. Specifically, the influence of the non-parabolicity on the interband transition has never been investigated.

\begin{figure*}
  \includegraphics{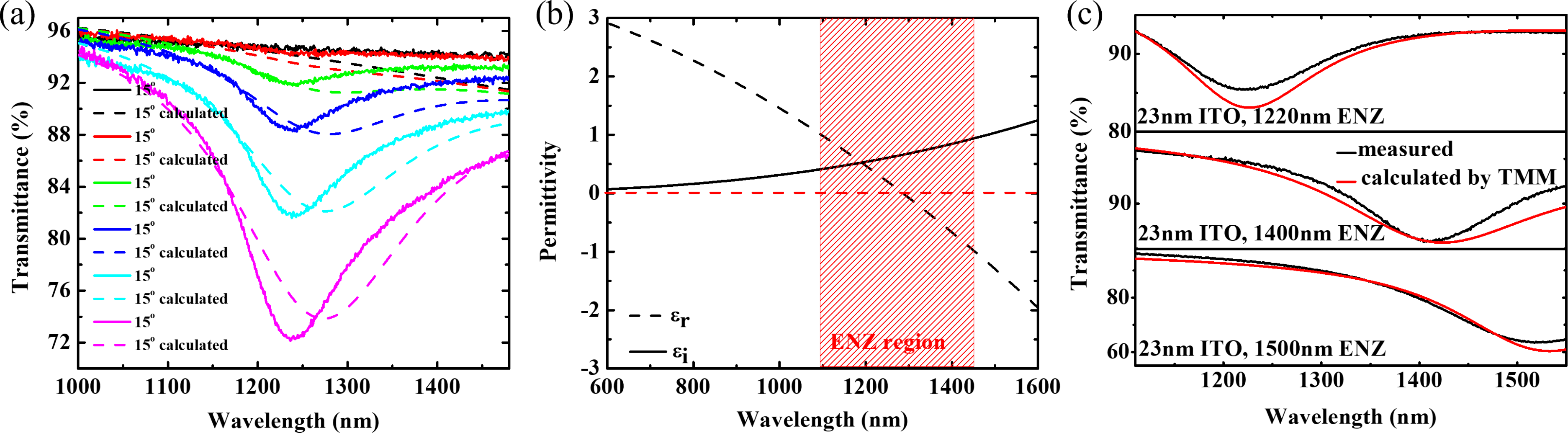}
  \caption{(a) Transmittance of the ITO film measured at different incident angles. The dashed curves show the fitted transmittance spectra obtained using the TMM and the Drude model. (b) Calculated permittivity of the commercial ITO film using the fitted Drude parameters. (c) Experimental (black line) and calculated (red line) 60$^{\circ}$  transmittance spectra of sputtered ITO films with different ENZ wavelengths.}
  \label{fig3}
\end{figure*}

Here, we explore the interaction and competition of the intraband and interband transitions in ITO films and report on a novel mechanism where the sign of the nonlinearity can be inverted. Our theoretical model predicts that by varying the non-parabolicity and the Fermi level of ITO, the ENZ resonance shift can be tuned to occur in opposite spectral directions. This optical switching of ENZ resonances gives rise to transient transmittance signals with highly variable spectral shapes. The theoretical predictions are validated using femtosecond (fs) pump-probe measurements with broadband probe beams on ITO thin films that have tunable ENZ wavelengths and demonstrate spectrally invertible nonlinearities. The non-parabolicity of the band structure, which contributes to a larger effective mass in the intraband transition, and also the Fermi energy, which determines the free carrier density, are shown to be two competing factors that jointly contribute to the invertible nonlinearity of ITO thin films.

Our work reveals the relationship between the large optical nonlinearity and the intrinsic dispersion relationships of ITO films in the ENZ region, which will enable design and development of TCO photonic materials and nonlinear devices. The ultrafast all-optical active control demonstrated here will aid applications such as all-optical modulators, integrated telecom circuits and other nonlinear photonic devices.

Fig.~\ref{fig3}a shows the linear optical properties of a 23 nm thick commercial ITO film on a glass cover slip. The oblique-angle transmittance measurements reveal a resonance peak at 1240 nm corresponding to the ENZ wavelength of ITO. The amplitude of the peak increases at larger incident angles, whereas normal incidence does not reveal an ENZ peak. Maxwell's boundary conditions show that
\begin{eqnarray}
  \epsilon_{0}E_{0\perp}=\epsilon_{1}E_{1\perp} \label{6}
\end{eqnarray}
where $\epsilon_0$ and $\epsilon_1$ are permittivity of air and the ITO film, respectively, and $E_{0\perp}$ and $E_{1\perp}$ are the normal components of electric field on both sides close to the ITO surface. Therefore, given a small $\epsilon_1$ in proximity to the ENZ region, the field normal to the ITO film would be enhanced with larger incident angles producing larger normal components of the field and hence a stronger field enhancement. To obtain the permittivity of the ITO thin film, a Drude model is adopted and the transmittance is calculated using transfer matrix method (TMM) with the fitted Drude parameters $\omega_p$ and $\gamma_p$. The dashed curves in Fig.~\ref{fig3}a show the calculation result using $\omega_p=1.9eV$ and $\gamma_p=0.17eV$ which show good agreement with the experimental measurements. The calculated dispersion of ITO with these Drude parameters is shown in Fig.~\ref{fig3}b. The real part of the permittivity switches sign at 1240 nm and defines the ENZ region with $\epsilon_i<1$ (shaded area in Fig.~\ref{fig3}b). To fabricate ITO thin films with different ENZ wavelengths, we use RF sputtering deposition on a glass substrate, followed by an annealing process (see Supplemental Material for fabrication details)~\cite{RN53}. The measured and fitted transmittance spectra of three samples at 60 degree excitation angle are shown in Fig.~\ref{fig3}c. All samples clearly show different ENZ resonance peaks in the NIR spectral region.

\begin{figure}
    \includegraphics{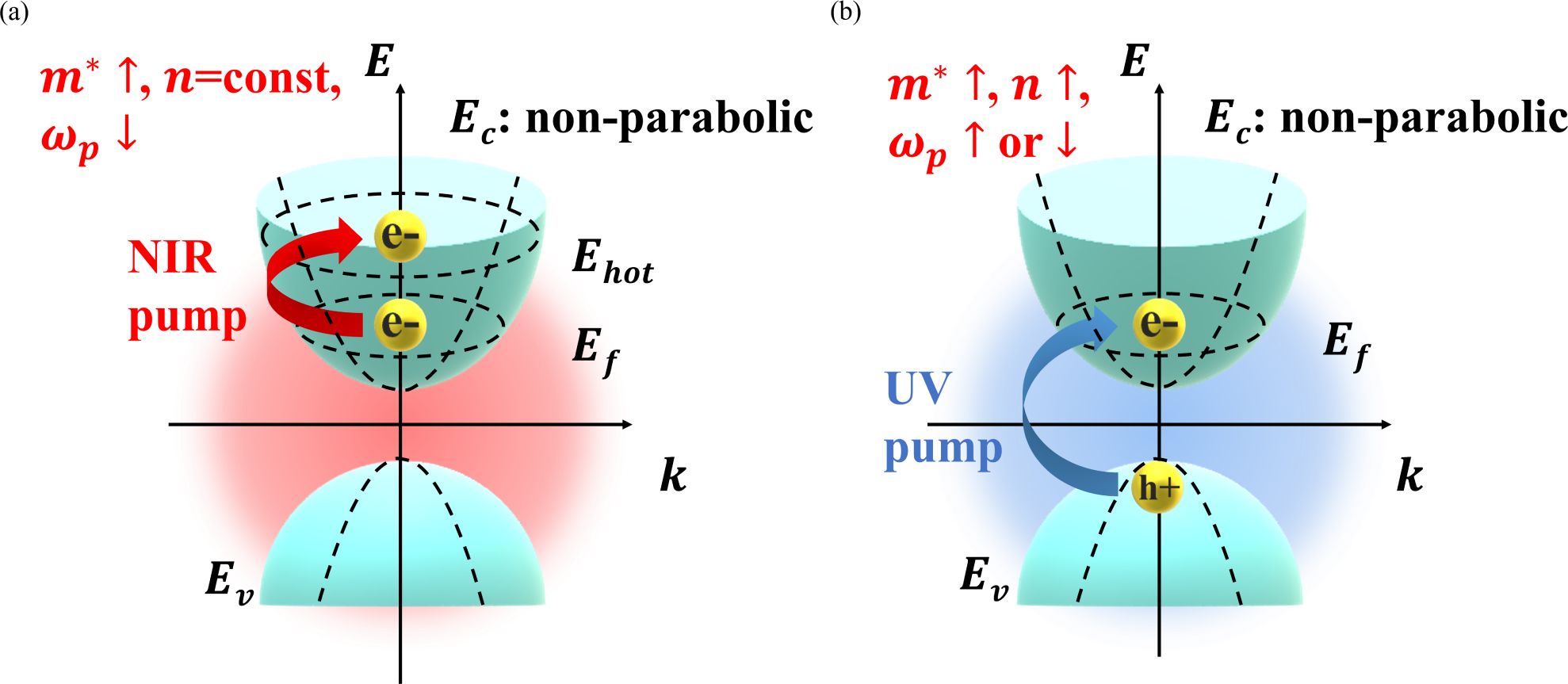}
    \caption{Schematic of the nonlinear optical responses in ITO thin films. (a) The intraband transition of conduction band electrons induces a larger effective mass due to the non-parabolic band structure and decreases the plasma frequency. $E_c$: conduction band, $E_f$: Fermi level, $E_v$: valence band, $E_{hot}$: hot electron energy level, dashed curves represent the parabolic band structure and the increase of the Fermi level. (b) The interband transition generates additional free carriers, increases the plasma frequency and induces a larger effective mass. The resulting nonlinearity depends on the competition between the two mechanisms and can be inverted by controlling the parabolicity and the Fermi level.}
    \label{fig1}
\end{figure}

Fig.~\ref{fig1} shows the schematic of transient optical processes. When the pump photon energy is smaller than the bandgap, the electrons in the conduction band undergo an intraband transition through free carrier absorption, as shown in Fig.~\ref{fig1}a. Before the pulse arrives, the electrons are in equilibrium and can be described as a room temperature Fermi distribution. Once the pump light is incident on the sample, the photon is absorbed and the electrons are excited to higher energy states to form a hot electron distribution. Due to the non-parabolicity of the conduction band, the effective mass of the electron sea is increased. This decreases the plasma frequency $\omega_p$
\begin{eqnarray}
  \omega_{p}=\sqrt{\frac{ne^2}{\epsilon_{0}m^{*}}} \label{1}
\end{eqnarray}
, where $n$ is the free electron density and $m^{*}$ is the effective mass of the conduction band, and red shifts the ENZ resonance. Since the band gap of ITO films is typically 3.5 eV-4.3 eV~\cite{yu2016indium}, this red shift happens when the pump wavelength is in the NIR region. On the other hand, as shown in Fig.~\ref{fig1}b, when the photon energy is greater than the bandgap, an interband transition occurs, where electrons in the valence band are promoted into the conduction band and the free carrier concentration $n$ is increased. Consequently, as revealed by Eq.~\ref{1}, $\omega_p$ increases and the ENZ resonance blue shifts accordingly. This happens for a UV pump where the photon energy is close to or above the band gap. However, with an increase of the free carrier density in the conduction band, the Fermi level is raised and the effective mass also increases due to the non-parabolicity. Thus, when the non-parabolicity of the material is large enough, then the effective mass increase dominates the free carrier density increase, and interband transition can cause an overall red shift of ENZ, inverting the nonlinearity. The total change and the shift direction of the plasma frequency and the nonlinearity will depend on the competing contributions of the the free carrier density in the equilibrium state (Fermi level) and the non-parabolicity of the conduction band structure.

\begin{figure*}
    \includegraphics{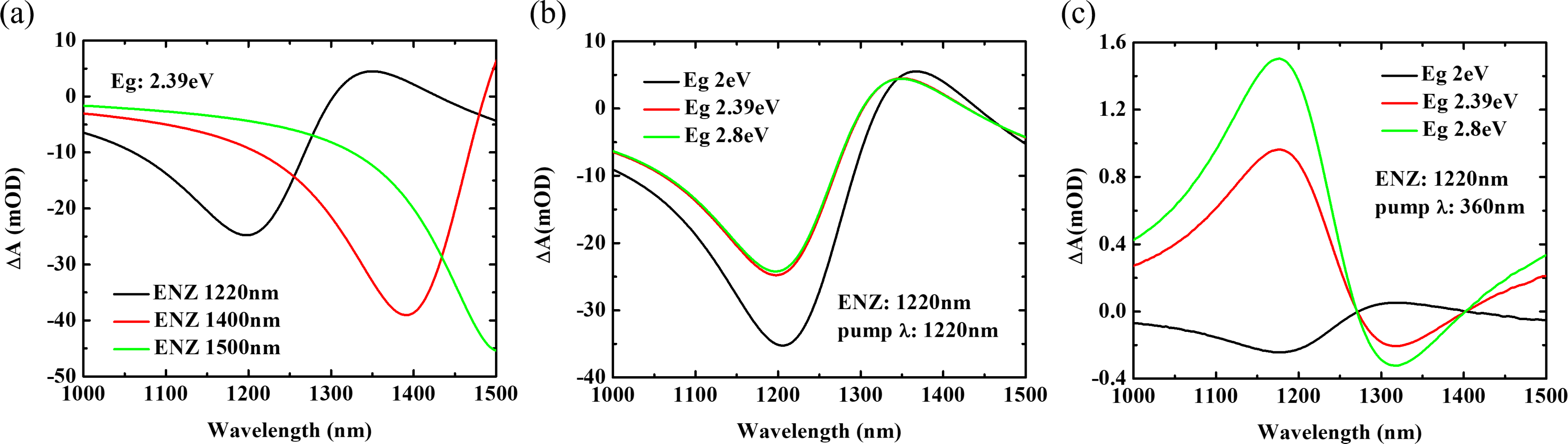}
    \caption{(a) Differential transmittance change ${\Delta}A$ with a broadband probe in NIR region calculated using the intraband transition model, where $E_g$ is 2.39 eV. Different colors represent different ENZ wavelengths. (b) Calculated ${\Delta}A$ for different values of $E_g$. The ENZ wavelength is 1220 nm. (c) Calculated ${\Delta}A$ spectra for various $E_g$ using the interband transition model. The ENZ wavelength is 1220 nm and the pump wavelength is 360 nm.}
    \label{fig2}
\end{figure*}

To evaluate the influence of both factors on the nonlinearity of ITO films, an analytical model is developed to describe the intraband and interband transitions. First, to account for non-parabolicity the band structure $E(k)$ is written as~\cite{kane1957band,cohen1961energy}:
\begin{equation}
  \frac{k^2\hbar^2}{2m^{*}}=E(k)+\frac{E^2}{E_g} \label{2}
\end{equation}
, where $1/E_g$ denotes the non-parabolicity. The plasma frequency $\omega_p$ is derived from the collisionless Boltzmann equation~\cite{RN52}. For the intraband transition, the free carrier density $n(\mu,T)$ is a constant and the electron energy density $U(\mu,T)$ is equal to the room-temperature energy density $U(\mu,300K)$ plus the absorbed pump photon energy density $dU$, where $\mu$ is the chemical potential and $T$ is the electron temperature (the detailed expressions for $\omega_p$, $n(\mu,T)$ and $U(\mu,300K)$ can be found in Ref.~\citenum{RN52}). Therefore, once the band structure parameters $m$, $E_g$ and the Drude parameters $\omega_p$, $\gamma_p$ are known, the nonlinearity caused by the intraband transition is fully determined. In Fig.~\ref{fig2}a and ~\ref{fig2}b, we theoretically explored the differential transmittance change ${\Delta}A$ near the ENZ region under the NIR pump, where intraband transition dominates the nonlinearity. ${\Delta}A$ is defined as:
\begin{eqnarray}
  {\Delta}A=-log(1+{\Delta}T/T_0) \label{3}
\end{eqnarray}
, where $T_0$ denotes the reference transmittance and ${\Delta}T$ is the dynamic change in transmittance after the pump. In Fig.~\ref{fig2}a, the ENZ wavelength is varied from 1220 nm to 1500 nm while the non-parabolicity $E_g$ and the effective mass are kept at 2.39 eV and $0.26m_0$, respectively, where $m_0$ is the electron mass at rest. At longer ENZ wavelengths that correspond to the lower free carrier density and the lower Fermi level, the maximum absolute value of ${\Delta}A$ is larger. This is because electrons at a lower Fermi level undergo a larger change in effective mass when the same amount of energy is absorbed which results in a larger nonlinearity. On the other hand, as shown in Fig.~\ref{fig2}b, ${\Delta}A$ is smaller with $E_g$ changing from 2 eV to 2.8 eV whereas the ENZ wavelength remains unchanged. Smaller $E_g$ represents larger non-parabolicity as shown in Eq.~\ref{2}, thus creating a larger change in the effective mass and a larger nonlinearity.

After individually studying the influence of the free carrier density and the non-parabolicity on the nonlinearity, we describe a cumulative phenomenon which involves both intraband and interband transitions under a UV pump. Under this condition, the free carrier density $n(\mu,T)$ is equal to the room-temperature carrier concentration $n(\mu,300K)$ plus the density of absorbed photons $dn$:
\begin{eqnarray}
   &n\left(\mu,300K\right)+dn=\frac{1}{\pi^2}\nonumber\\
   &\times\int_{0}^{\infty}{dE\left(\frac{2m}{\hbar^2}\left(E+\frac{E^2}{E_g}\right)\right)^\frac{1}{2}\frac{m}{\hbar^2}\left(1+\frac{2E}{E_g}\right)f_0\left(\mu,T\right)} \label{4}
\end{eqnarray}
, where $f_0(\mu,T)$ represents the Fermi-Dirac distribution. $dn$ is given by:
\begin{eqnarray}
    dn=\frac{FAcos\theta}{\pi{r^2}h{E_{photon}}} \label{5}
\end{eqnarray}
, where $F$ is the energy per pulse, $A$ is the absorption, $\theta$ is the incident angle, $r$ is the pump beam radius, $h$ is the thickness of the film and $E_{photon}$ is the photon energy. The calculated ${\Delta}A$ spectra under the UV pump is plotted in Fig.~\ref{fig2}c, where the ENZ wavelength is 1220 nm and $E_g$ is varied from 2 eV to 2.8 eV. Interestingly, unlike in Fig.~\ref{fig2}a and \ref{fig2}b where the NIR pump gives the same spectral shape of ${\Delta}A$, the spectral shape under the UV excitation in Fig.~\ref{fig2}c depends on the relative strength between the free carrier density and the non-parabolicity. If $E_g$ is relatively small (Fig.~\ref{fig2}c, black curve) the hot electron-induced Fermi level shift has a stronger contribution than the increase in carrier density due to band-to-band transition and causes a red shift of the plasma frequency $\omega_p$. Conversely, an overall blue shift of $\omega_p$ occurs when the hot electron contribution is relatively lower (Fig.~\ref{fig2}c, red and green curves). To the best of our knowledge, this is the first observation of spectral inversion of nonlinearity based on the competing mechanisms of the free carrier density changes and the non-parabolicity.

\begin{figure}
  \includegraphics{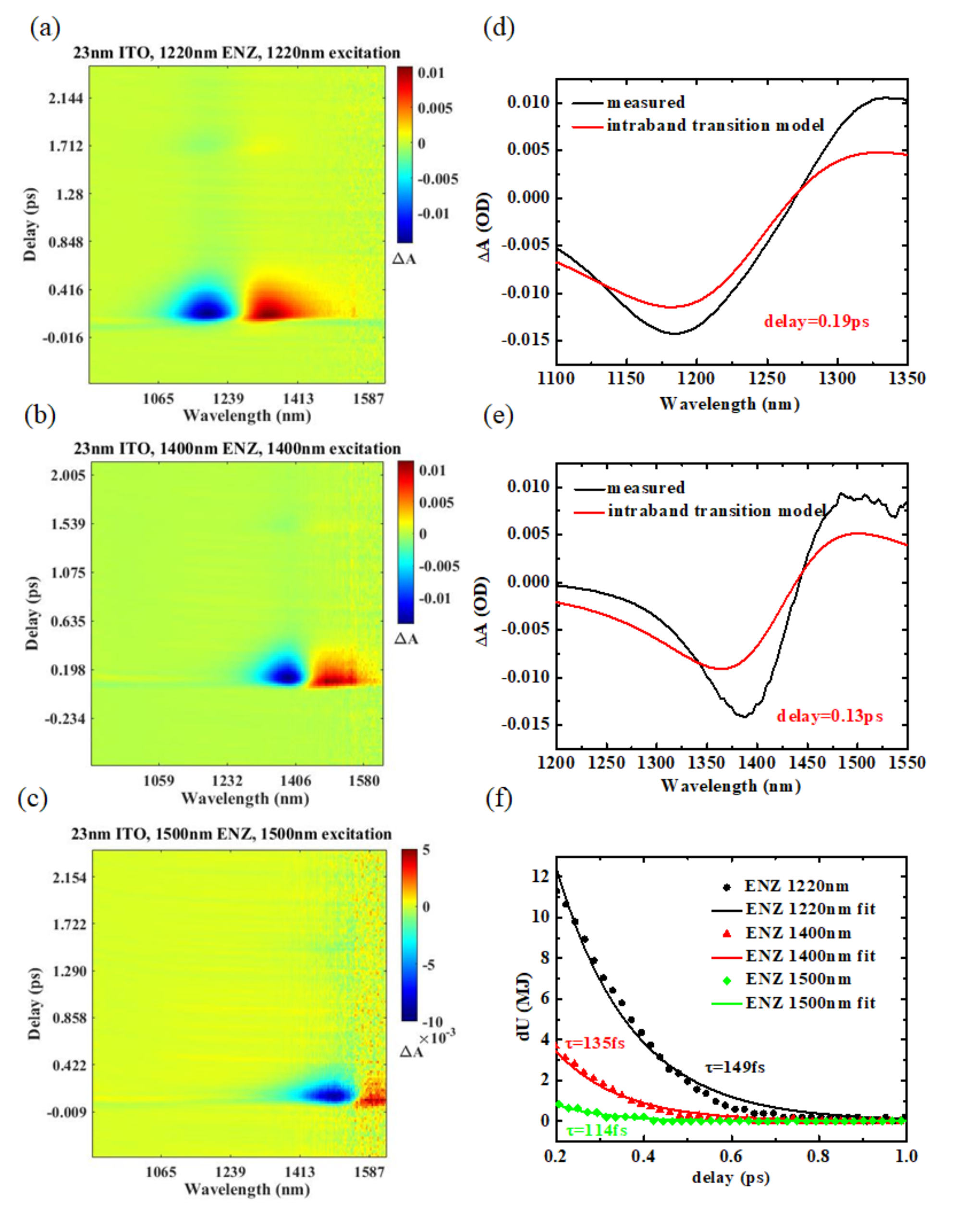}
  \caption{(a)-(c) Time-resolved transmittance spectral change of ITO films. The pump wavelength coincides at the ENZ wavelength at (a) 1220 nm (b) 1400 nm and (c) 1500 nm. (d), (e) ${\Delta}A$ spectra at the beginning of the relaxation process for ENZ wavelengths at (d) 1220 nm and (e) 1400 nm. The black curves are extracted from the experimental data with the delay time at (d) 0.19 ps and (e) 0.13 ps. The red curves fit the plasma frequency $omega_p$ after pump. (f) kinetic decay process of the absorbed energy $dU$. The pump wavelengths are in the NIR region. The squares show experimental data and solid lines are exponential fits.}
  \label{fig4}
\end{figure}

Next, we turn our attention to optical pump-probe measurements to elucidate the impact of the Fermi level and the non-parabolicity on ITO nonlinearities. In a set of transient ultrafast measurements we use a narrowband fs pump laser pulse to induce excitations in ITO films and a broadband fs probe pulse at various delay times to measure the time-resolved transmittance spectra (see Supplemental Material for details of the experimental setup). Fig.~\ref{fig4}a shows the ${\Delta}A$ spectral map measured on a ITO film with a ENZ resonance at 1220 nm. A pump wavelength that coincides with the ENZ wavelength optimizes the absorption and enhances the nonlinear response. An ultrafast red-shift of the ENZ resonance is observed, indicating the increase in the electron effective mass induced by intraband transitions. A maximum of nonlinear response of ${\Delta}A=100mOD$ is obtained at a pump fluence of $1000{\mu}J/cm^2$, as shown in Fig.~S1, corresponding to ${\Delta}T/T=35\%$, which is comparable to values in the literature~\cite{RN10,RN30,RN52}. Additional ${\Delta}A$ spectral maps measured on ITO films with different ENZ wavelengths and pumped at their resonance peaks are plotted in Fig.~\ref{fig4}b and Fig.~\ref{fig4}c. In Fig.~\ref{fig4}d, the spectrum at 0.19 ps from Fig.~\ref{fig4}a corresponding to the time delay immediately after the pump pulse absorbtion is fitted using the TMM to get the plasma frequency $\omega_{p,pump}$. Using the fitted $\omega_{p,pump}$ and unpumped $\omega_{p,unpump}$, the effective mass $m^{*}$ and the non-parabolicity $E_g$ can be determined from the intraband transition model. The theoretical spectrum (Fig.~\ref{fig4}d, red curve) is in good agreement with the experimental data. The same calculation is conducted on other samples with ENZ wavelengths at 1400 nm and 1500 nm, as shown in Fig.~\ref{fig4}e and Fig. S2. All the theoretical data match well with the measurements, confirming the validity of our model. The values for all Drude and band structure parameters are listed in Table S2.

After absorbing the pump photon energy and creating a hot electron distribution, the carriers cool down through the electron-phonon coupling and return to the equilibrium state on a sub-ps time scale~\cite{boyd,del2000nonequilibrium}. The dissipation of the excitation energy density $dU$ as a function of time, is plotted in Fig.~\ref{fig4}f and represents the dynamics of the relaxation process. The dashed lines are exponential fits to extract the relaxation times. These values are typically $\sim$100 fs and agree well with ultrafast intraband transition dynamics. The increasing discrepancy between the fitted data and measurements at shorter delay times can be attributed to the electron temperature-dependent heat capacity of electron gas~\cite{boyd}.

\begin{figure}
  \includegraphics{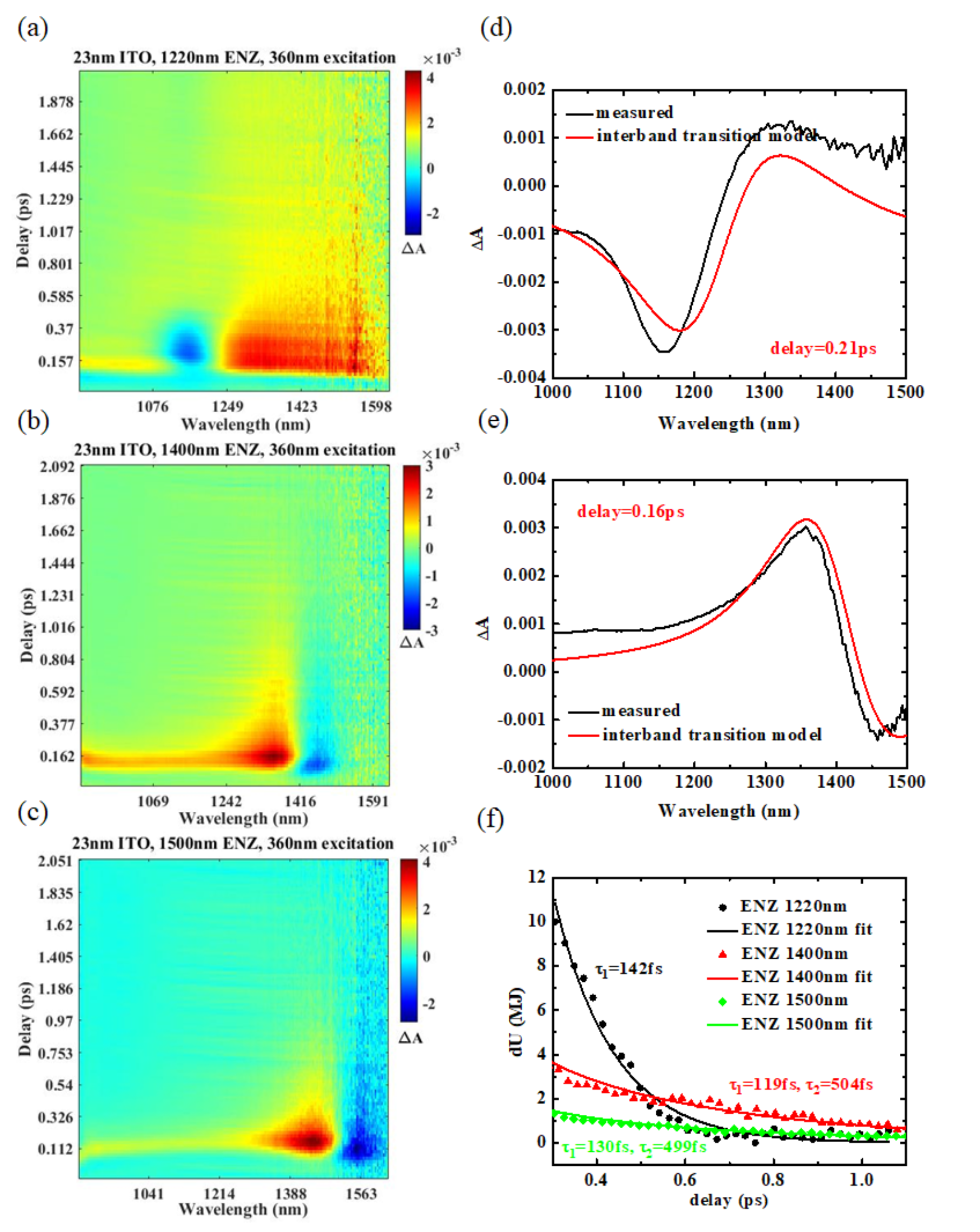}
  \caption{(a)-(c) Time-resolved transmittance spectral change of ITO films in pump-probe measurements. The pump wavelength is at 360 nm. The ENZ wavelength is at (a) 1220 nm, (b) 1400 nm and, (c) 1500 nm. (d),(e) ${\Delta}A$ spectra at the beginning of the relaxation process for ENZ wavelengths at (d) 1220 nm and (e) 1400 nm. The black curves are extracted from the experimental data with the delay time at (d) 0.21 ps and (e) 0.16 ps. The red curves fit the plasma frequency $\omega_p$ after pump. (f) kinetic decay process of the absorbed energy $dU$. The pump wavelength is 360 nm. The squares show experimental data and solid lines are exponential fits.}
\label{fig5}
\end{figure}

Next, a UV pulse centered at 360 nm is used as the pump source to introduce interband transitions in ITO films. The time-resolved transmittance maps measured on ITO films with various ENZ wavelengths are shown in Fig.~\ref{fig5}a-c.
The photon energy at 360 nm corresponds to the band gap of these ITO films as shown in Fig. S4, which shows the pump-wavelength dependent maximum of ${\Delta}A$. The Burstein-Moss bandgap shift~\cite{burstein1954anomalous,meng1998properties} has been shown to be negligible, thus it is a reasonable assumption that the same amount of electrons in the valence band are excited to the conduction band in all samples. However, different nonlinear behavior is clearly observed as a result of the competing mechanisms of  inter- band and intraband absorption. The ITO film with ENZ wavelength at 1220 nm has the largest non- parabolicity (see Table S2), therefore the contribution of intraband transition to the plasma frequency is larger than the contribution of interband transition. Finally, as shown in Fig.~\ref{fig5}a, the plasma frequency is red-shifted which is similar to the behavior under the NIR pump. Conversely, films with longer ENZ wavelengths have smaller non-parabolicity, which is not able to balance the shift induced by the free carrier density increase induced by interband transitions. Therefore, the plasma frequency is blue-shifted as shown in Fig.~\ref{fig5}b and Fig.~\ref{fig5}c. These results agree well with the theoretical calculations shown in Fig.~\ref{fig2}c. Moreover, in Fig.~\ref{fig5}d, Fig.~\ref{fig5}e and Fig. S3, the experimental spectra corresponding to the peak of ${\Delta}A$ are plotted along with the calculation for $\omega_{p,pump}$. The same parameters in Table S2 are used to fit the spectra with the interband transition model (Fig.~\ref{fig5}d, Fig.~\ref{fig5}e and Fig. S3, red curves). The calculation results are verified by the measurements, indicating the validity of our theoretical predictions.

Finally, the dynamics of the relaxation process is investigated. This process involves thermal dissipation, where the electron temperature decreases through electron-phonon interactions and recombination, including the radiative and the trap-assisted recombination channels~\cite{sze2021physics}. Generally the recombination is slower than the thermal dissipation process. As shown in Fig.~\ref{fig5}f (black curve), the decay of $dU$ in the ITO film with the ENZ wavelength at 1220 nm shows a characteristic recovery time of 142 fs. This agrees with the typical thermal dissipation time revealed in Fig.~\ref{fig4}f, indicating the dominance of the intraband transition. On the other hand, red and green curves in Fig.~\ref{fig5}f include a slower decay process with $\tau_{2}$ around 500 fs in addition to the faster decay $\tau_{1}$ attributed to thermal dissipation. This longer relaxation rate may correspond to the recombination, which again indicates that interband transitions plays a dominant role in ITO films with the ENZ at 1400 nm and 1500 nm.

In conclusion, we have explained the role of various mechanisms responsible for ultrafast nonlinear properties of ITO thin films near their ENZ wavelengths using a comprehensive study that includes analytical models verified by experimental measurements. We have explored the interplay between the intraband transition, which increases the electron effective mass due to the non-parabolic band structure, and the interband transition, which generates larger free carrier concentrations. The combination of the two processes under the above-bandgap pump results in different nonlinear behaviors depending on the relative contributions between the non-parabolicity and the Fermi level shift. An interband transition model is used to describe these competing mechanisms and predict the cumulative shift direction of the plasma frequency. Transient transmittance measurements agree well with the theoretical predictions demonstrating the spectrally invertible nonlinearity. Moreover, the Drude parameters and band structure parameters extracted from the fits show clear trends that are related to the material properties of ITO films. Our findings may guide the design and fabrication of TCO photonic platforms based on the tunability of nonlinear optical interactions.

\begin{acknowledgments}
This work was supported by UDRI with the Air Force contract no. FA8651-20-D-0003. MSA, SUD, and JWA are thankful for the funding support through AFOSR Lab Tasks 22RWCOR002. HH acknowledges support from the Department of Energy (DE-SC0020101).
\end{acknowledgments}

\end{document}